\documentclass[aps,prl,twocolumn,nofootinbib,showpacs,floatfix]{revtex4-1}

\usepackage{graphicx}
\usepackage{amsmath}
\usepackage{amssymb}
\usepackage{color}
\usepackage{txfonts}
\usepackage{verbatim}

\begin{document}

\title{Persistent Astrometric Deflections from Gravitational-Wave Memory}
\author{Dustin R. Madison}
\email{dustin.madison@mail.wvu.edu}
\affiliation{Department of Physics and Astronomy, West Virginia University, P.O. Box 6315, Morgantown, West Virginia 26506, USA}
\affiliation{Center for Gravitational Waves and Cosmology, West Virginia University, Chestnut Ridge Research Building, Morgantown, West Virginia 26505, USA}

\date{\today}
\begin{abstract}
Gravitational waves (GWs) produce small distortions in the observable distribution of stars in the sky. We describe the characteristic pattern of astrometric deflections created by a specific gravitational waveform called a burst with memory. Memory is a permanent, residual distortion of space left in the wake of GWs. We demonstrate that the astrometric effects of GW memory are qualitatively distinct from those of more broadly considered, oscillatory GWs---distinct in ways with potentially far-reaching observational implications. We discuss some such implications pertaining to the random-walk development of memory-induced deflection signatures over cosmological time spans and how those may influence observations of the cosmic microwave background.
\end{abstract}
\maketitle

\paragraph{Introduction.---}
Gravitational wave (GW) memory is an anticipated component of most gravitational waveforms, especially bursts. GW bursts with memory cause relative displacements within systems of free-falling masses that last indefinitely after the GW burst has passed through the system. For sources of GWs like binary black holes, memory is sourced by GW emission and how the waves influence the evolution of the stress-energy tensor. In other words, memory is a GW effect generated by GW emission---a deeply nonlinear probe of extreme gravity \citep{bt87,c91,t92,bd92,f09,f10,ttl+18}. Furthermore, detecting memory may speak to the structure of the Universe as a whole and the nature of gravity on cosmological scales. Memory can be viewed as a direct observational consequence of the infinite number of symmetries and conservation laws in general relativity and may be tied to the resolution of the black hole information paradox, an important puzzle of theoretical physics \citep{sz16,hps16,hps17,hiw17,fn17}. 

Memory generated by the mergers of supermassive black hole binaries could be detected by pulsar timing arrays (PTAs) \citep{vl10,cj12,mcc14,whc+15}, but the rates of detectable mergers are expected to be so low that it is unlikely for one to occur during these decades-long projects. A memory event of strain amplitude $10^{-14}$, a nearly detectable amplitude for current PTAs, may occur only once per $10^6$ or $10^7$ years \citep{aab+20,isb+19}. Pendula in ground-based GW detectors provide restoring forces that erase permanent memory-induced displacements, but it is possible that transient influences from memory can be inferred in an ensemble of signals once thousands of GW events are detected \citep{ltl+16,htl+20}. Other probes of memory will prove indispensable for investigating this important phenomenon.    

Ground-based detectors and PTAs look for perturbations to light travel times caused by GWs. But GWs also cause deflections in the apparent positions of distant sources of light \citep{ksg+99,bf11}. The prospects for detecting GWs through astrometric deflections were initially deemed pessimistic \citep{s09}, but the success of the astrometric mission {\sl Gaia} \citep{Gaia18} in measuring submilliarcsecond precision positions for over a billion sources has led some to reconsider this pessimism. \citet{mml+17} demonstrated that {\sl Gaia} could have sensitivity to individually resolvable sources of GWs rivaling that of PTAs. Recent results from the {\sl Event Horizon Telescope} have highlighted the advancing capabilities of very-long-baseline interferometry (VLBI) \cite{EHT19}. \citet{dtp18}, supplementing {\sl Gaia} results with VLBI measurements, derived constraints on the energy density in a stochastic GW background in and below the range of frequencies accessible with PTAs.

Motivated by theoretical interest in GW memory, the astrometric capabilites of instruments like {\sl Gaia}, the {\sl Event Horizon Telescope}, and their successors, and the difficulties that PTAs and ground-based GW detectors will face in detecting memory, we investigate the astrometric signature of GW bursts with memory. We demonstrate that memory is distinct from other types of GWs in this forum, yielding enticing prospects for detection and future inquiry.     

\paragraph{Deflections from a Planar GW.---}
Consider a planar GW traveling in direction ${\bf \hat{p}}$ with polarization $\varepsilon_{ij}$: 
\begin{equation}
\label{eq:planewave}
h_{ij}(t,{\bf x})=\varepsilon_{ij}h(t-{\bf\hat{p}}\cdot{\bf x}).
\end{equation}
To describe the astrometric deflections from this GW, we adopt the formalism of \citet{bf11}. For a light source in direction ${\bf \hat{n}}$, the unperturbed worldline of light arriving at the origin at time $t$ is
\begin{equation}
x_0^\alpha(\lambda)=(\lambda+ct,-\lambda{\bf \hat{n}}). 
\end{equation}
The parameter $\lambda$ varies from $-d_s$ to $0$ where $d_s$ is the distance between the observer and light source when the light was emitted. The deflection of the light source is given by Eq.~(35) from \citet{bf11}:
\begin{eqnarray}
\label{eq:deflect_general}
\delta n^i(t)&=&Q^{ik}\hat{n}^j\left\{-\frac{1}{2}h_{jk}(t,{\bf 0})+\frac{\hat{p}_k\hat{n}_l}{2(1+{\bf \hat{p}}\cdot{\bf \hat{n}})}h_{jl}(t,{\bf 0})\right.\nonumber\\
&&\left.+\frac{1}{d_s}\int_{-d_s}^0d\lambda\left[h_{jk}(\lambda)-\frac{\hat{p}_k\hat{n}_l}{2(1+{\bf \hat{p}}\cdot{\bf \hat{n}})}h_{jl}(\lambda)\right]\right\},\nonumber\\
\end{eqnarray}
where $Q^{ik}=(\delta^{ik}-\hat{n}^i\hat{n}^k)$ and $\delta^{ik}$ is the Kronecker delta. The integral is over the unperturbed photon worldline. We used the shorthand $h_{ij}(\lambda)$ to mean the GW field along the integration path, i.e.,
\begin{eqnarray}
h_{ij}(\lambda)=\varepsilon_{ij}h\left(t+\frac{\lambda}{c}(1+{\bf \hat{p}}\cdot{\bf \hat{n}})\right).
\end{eqnarray}

Equation (\ref{eq:deflect_general}) can be reexpressed as  
\begin{eqnarray}
\label{eq:deflect_specialized}
\delta n^i(t)&=&{\cal V}^i_{\oplus}h(t)-{\cal V}^i_{\bigstar}\frac{H(t)-H(t_L)}{t-t_L},
\end{eqnarray}
where 
\begin{eqnarray}
\label{eq:delay_time}
t_L&=&t-\frac{d_s}{c}(1+{\bf \hat{p}}\cdot{\bf \hat{n}}),\\
{\cal V}^i_{\oplus}&=&\frac{\hat{p}^i+\hat{n}^i}{2(1+{\bf \hat{p}}\cdot{\bf \hat{n}})}\hat{n}^j\hat{n}^k\varepsilon_{jk}-\frac{1}{2}\hat{n}^j\varepsilon_j^{~i},\\
{\cal V}^i_{\bigstar}&=&{\cal V}^i_\oplus-\frac{1}{2}Q^{ik}\hat{n}^j\varepsilon_{jk},
\end{eqnarray}
and $H$ is the antiderivative of $h$. The term in Equation (\ref{eq:deflect_specialized}) proportional to ${\cal V}^i_\oplus$ was first derived by \citet{pgb+96} and was the starting point of the analysis by \citet{mml+17} who referred to it as the ``Earth" term in analogy to PTA parlance and because it only depends on the GW strain at the location of Earth. \citet{mml+17} did not give an explicit mathematical expression for the other term in Equation~(\ref{eq:deflect_specialized}), but they dubbed it the ``star" term and explained why they could justifiably ignore it. For a planar GW of reduced wavelength $\lambdabar$, $h(t)=h_0\cos{(ct/\lambdabar+\varphi)}$ for some phase $\varphi$. The antiderivative is $H(t)=h_0(\lambdabar/c)\sin{(ct/\lambdabar+\varphi)}$. With $t-t_L=(d_s/c)(1+{\bf \hat{p}}\cdot{\bf \hat{n}})$ in the denominator of the star term, it is overall proportional to $h_0\lambdabar/d_s$. As long as $\lambdabar\ll d_s$, the star term is negligible compared to the Earth term. 

The redshift from a GW, which is relevant for PTAs, splits into Earth and pulsar terms [see, e.g., Eq. (29) from Book \& Flanagan \cite{bf11}]. The redshift Earth term depends on the strain at Earth, while the redshift pulsar term depends on the strain at the pulsar at some time in the past (the same delay as in our expression for $t_L$). The Earth and pulsar terms of the redshift share the same geometric prefactor. The splitting of the astrometric deflection in our Eq. (\ref{eq:deflect_specialized}) is very different from the splitting in the redshift. The deflection Earth term does depend on the strain at Earth, but it has a different geometric dependence than the star term, already differentiating the deflection and redshift signatures. Furthermore, the star term depends on the integral of the strain along the unperturbed worldline, hence the appearance of the antiderivative in Eq.~(\ref{eq:deflect_specialized}). The appearance of $t-t_L~\propto~d_s$ in the star term's denominator further distinguishes it from the pulsar term and renders it negligible in most situations. When memory is considered, however, the star term produces non-negligible effects.

\begin{figure}[b]
    \centering
    \includegraphics[scale=.55,trim=0 0 0 1cm ]{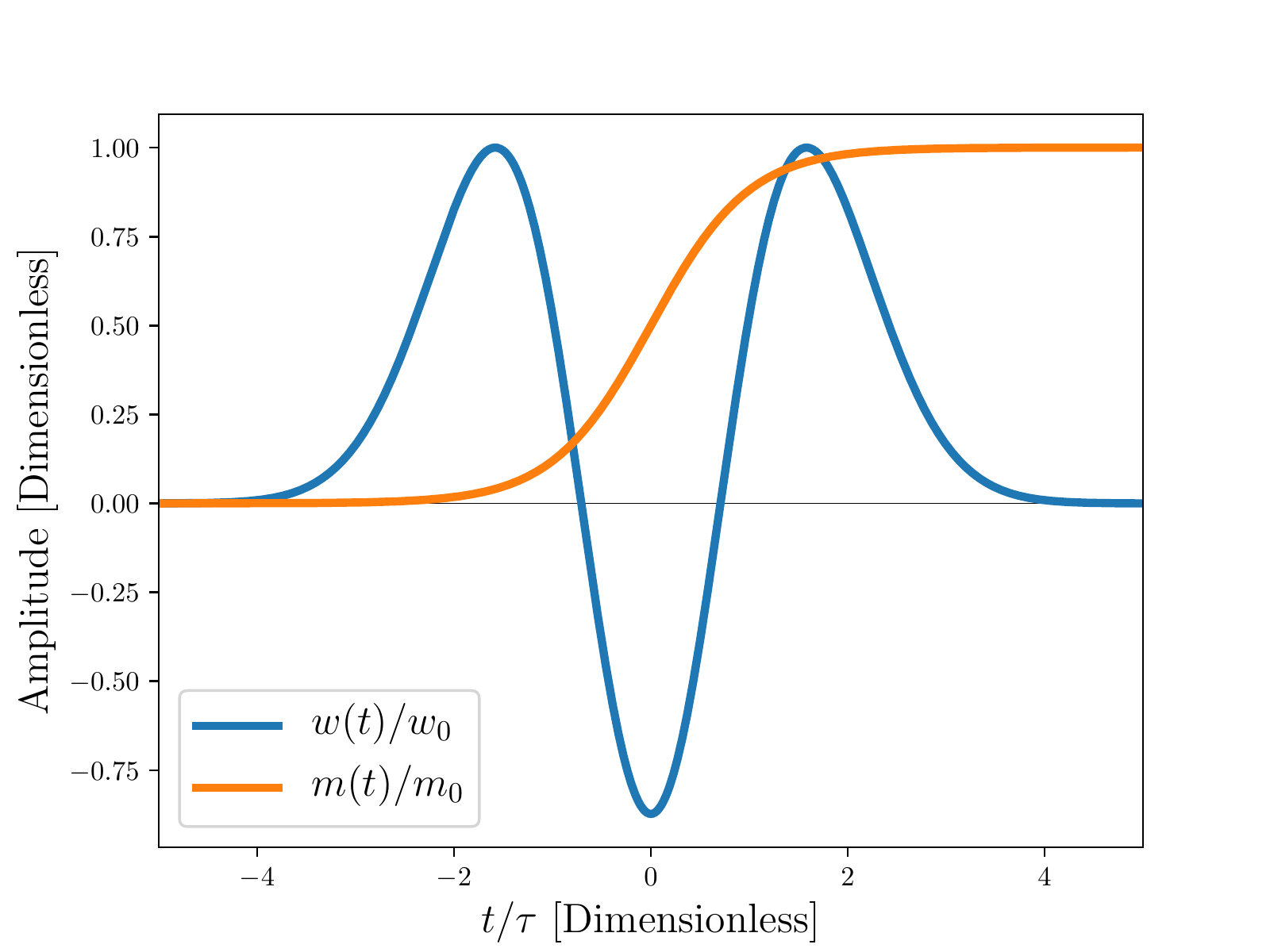}
    \caption{We consider astrometric deflections from waveletlike GWs (blue), memorylike GWs (orange), and superpositions thereof.}
    \label{fig:waveforms}
\end{figure}

\paragraph{GW Bursts with Memory.---}
We consider GWs with time dependence $h(t)=m(t)+w(t)$, where
\begin{eqnarray}
\label{eq:memory}
m(t)&=&\frac{m_0}{2}\left[\tanh{\left(\frac{t}{\tau_m}\right)}+1\right],~~{\rm and}\\
\label{eq:wavelet}
w(t)&=&\frac{w_0}{4}e^{5/4}\left[2\left(\frac{t}{\tau_w}\right)^2-1\right]e^{-t^2/2\tau_w^2}.
\end{eqnarray}
These curves are shown in Fig. \ref{fig:waveforms}. Memory is modeled by $m(t)$, growing from zero to some final value $m_0$, over a timescale $\tau_m$ and qualitatively resembling the memory waveforms produced by Favata to describe a compact binary coalescence \cite{f09,f10}. An oscillatory, waveletlike component of the GW, $w(t)$, has maximum absolute amplitude $w_0$ and is temporally localized to an interval of duration $\tau_w$. For future use, we define the total area under $w(t)$ as $W_0$.  

This simple signal model is broadly descriptive of essentially any bursting emission of GWs: temporally localized oscillations paired with monotonically growing memory. Taking binary black hole mergers as realistic examples, the characteristic rise time for the memory $\tau_m$ is approximately the light travel time around the postmerger event horizon, scaling linearly with the total mass and equal to approximately one day for a billion solar mass merger \cite{mcc14}; the timescales $\tau_m$ and $\tau_w$ are approximately equal as the memory grows rapidly through the last cycles of inspiral. We take $\tau_w=\tau_m$ and refer to both as $\tau$. Though details like the mass ratio and the relative orientation of black hole spins can have non-negligible effects, broadly speaking, $m_0\approx0.1~w_0$ in physical systems \citep{t92,f09}. If one wanted, $w(t)$ could be replaced by the final cycles of a chirping waveform from a binary inspiral. As we demonstrate, the details of $w(t)$ are not important for describing the long-term behavior of the deflection. Memory is key.

\begin{figure}[b]
    \centering
    \includegraphics[scale=.55,trim=0 0 0 1cm]{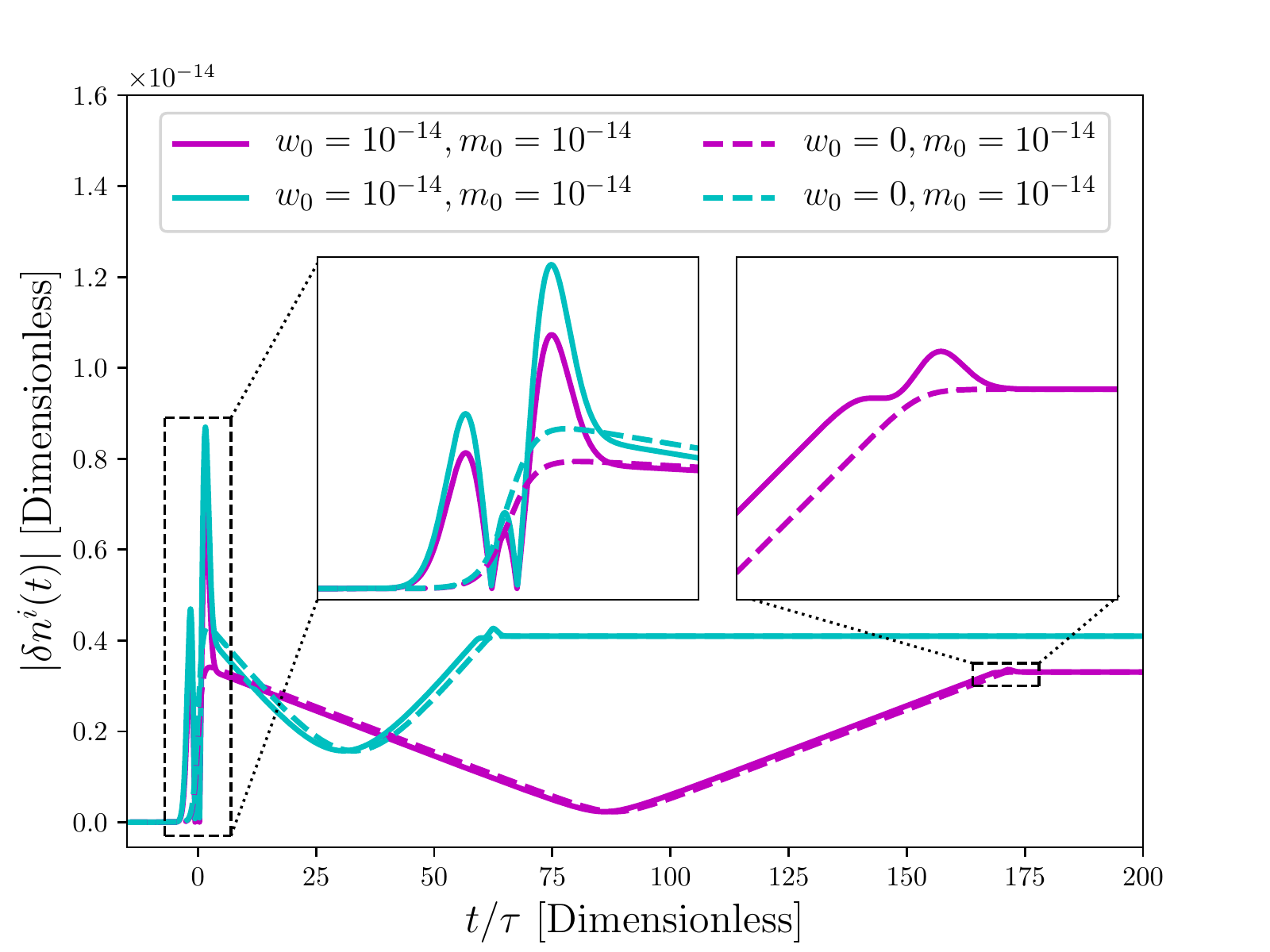}
    \caption{The norm of the deflection vector, $\delta n^i$, as a function of time for a GW propagating in the positive ${\bf \hat{z}}$ direction. The deflection vector is given by Eq. (\ref{eq:deflect_specialized}) and we have specifically analyzed GW waveforms that are combinations of Eqs. (\ref{eq:memory}) and (\ref{eq:wavelet}) (scales as dictated by the legend). The two colors correspond to sources of light in different directions (see the color-coded ``star" markers in Fig. \ref{fig:displacements}). The solid curves describe the influence of an equal-weight superposition of a wavelet signal and a memory signal. The dashed curves describe the influence of the memory alone. We set the distance of the light sources to $c\tau\times10^2$. The left-hand inset details early times near $t=0$ when the GW waveform passes over Earth. The right-hand inset details times near $t_L$ for the sky direction associated with the magenta star in Fig. \ref{fig:displacements}. A similar inset detailing the corresponding part of the cyan curve would look almost identical.}
    \label{fig:deflection_norms}
\end{figure}

\begin{figure}[b]
    \centering
    \includegraphics[scale=.4,trim=2cm 0 0 1cm]{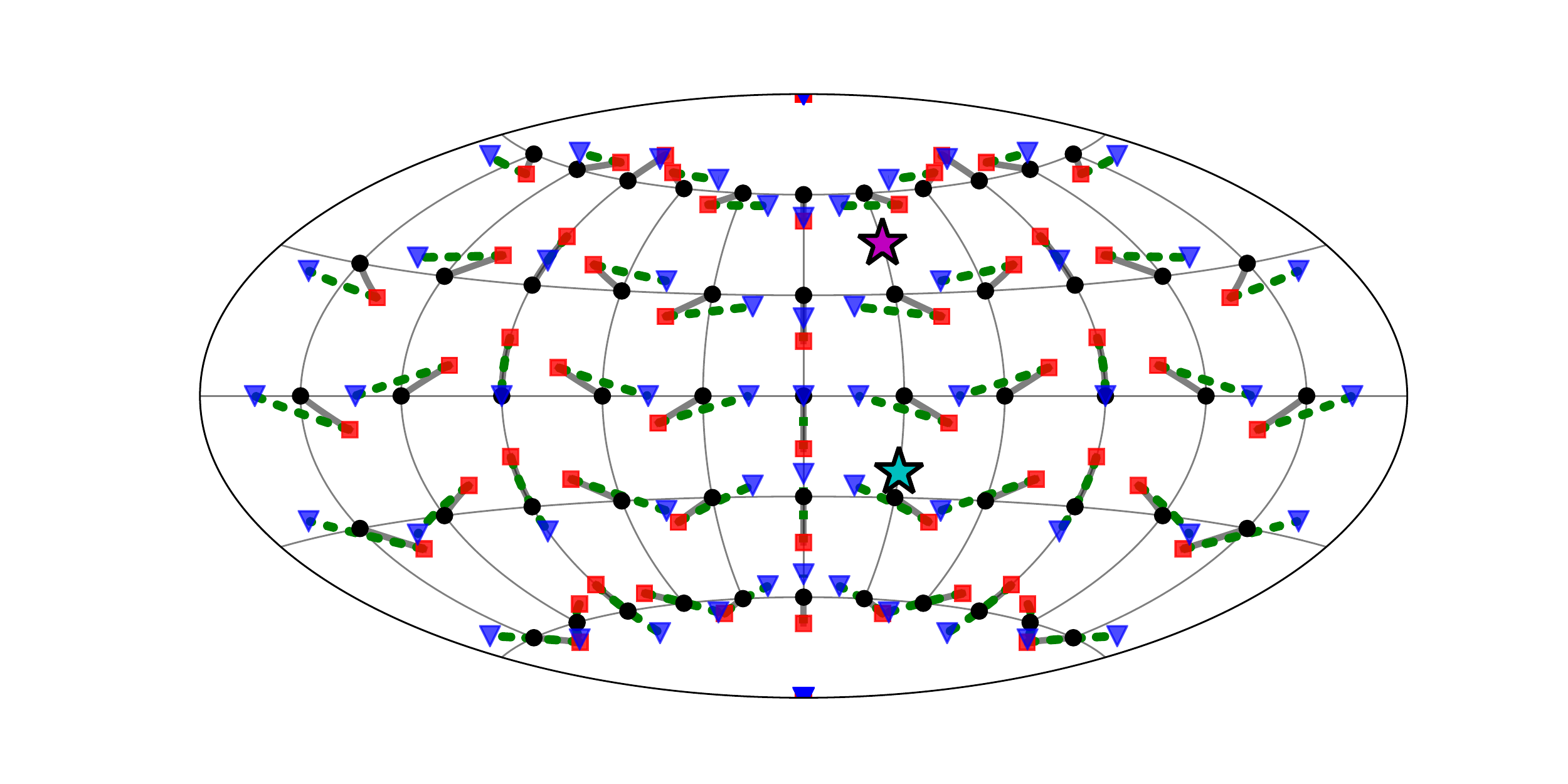}
    \caption{The prompt and permanent deflection patterns caused by GW memory. For the sake of visualization, we inflated the magnitude of the effect to unrealistically large levels. Over a timescale $\tau$, the time it takes for memory to grow from zero to its final value, light sources originally at the locations of the black dots are deflected to the positions of the red squares as the wavefront passes over Earth. Over a long timescale $(d_s/c)(1+{\bf\hat{z}}\cdot{\bf\hat{n}})$, light sources drift along the dotted green paths, eventually arriving at the positions of the blue triangles.}
    \label{fig:displacements}
\end{figure}

In Fig. \ref{fig:deflection_norms}, we show the norm of the deflection vector $|\delta n^i|$ caused by a GW with propagation direction ${\bf\hat{p}=\hat{z}}$. We show the full span of relevant times, ranging between $t=0$ and $t=2d_s/c$, the maximum value of $t-t_L$. We fixed $d_s$ at $c\tau\times 10^2$. As we said, even for a billion solar mass binary black hole merger, $\tau$ is only approximately a day, so this choice of $d_s$ is exceedingly small compared to astronomically relevant distances. More realistic distances with $\tau=1$ day would be $d_s\approx c\tau\times10^7$ for sources of light still within our Galaxy. However, this small distance allows us to clearly display the entire evolution of $|\delta n^i(t)|$ in Fig. \ref{fig:deflection_norms}. The two colors of curves describe the deflections for light sources with the same longitude but different latitudes, magenta in the northern hemisphere and cyan in the southern (the positions of the color-coded ``star" markers in Fig. \ref{fig:displacements}). With ${\bf\hat{p}=\hat{z}}$, a star's latitude determines the value of ${\bf \hat{p}\cdot\hat{n}}$ that enters the expressions for ${\cal V}^i_\oplus$, ${\cal V}^i_\bigstar$, and $t_L$. A star's longitude matters only through its relation to the polarization of the GW, which we have taken as ``$+$',' aligned with the $x$ and $y$ axes.  The solid curves describe the case where $m_0=w_0=10^{-14}$, an equal superposition of the wavelet and memory signals. The dashed curves show the influence of the memory component of the signal alone.

The wavelet component of the burst generates transient fluctuations through the Earth term with amplitude proportional to $w_0$ (detailed in the left-hand inset of Fig. \ref{fig:deflection_norms}). As the wavelet completes its passage over Earth, it leaves, through the star term, a long-lived, constant, but small deflection proportional to $cW_0/d_s$, where $W_0$ is the total area under $w(t)$. This is why the dashed and solid curves do not perfectly overlap after the initial transient fluctuations. After a time $t_L$, the star term generates low-amplitude fluctuations that negate the small, long-lived deflection from the wavelet. These ripples can be seen in the right-hand inset of Fig. \ref{fig:deflection_norms}. An inset focusing on the corresponding region of the cyan curve would look nearly identical. If we had chosen a larger, more astronomically realistic value for $d_s$, the offset between the dashed and solid curves following the transient fluctuations and the small ripples depicted in the right-hand inset would be imperceptible as the star term would be more significantly suppressed for short wavelength signals. 

Besides the initial transient fluctuations and the small, long-lived deflection proportional to $cW_0/d_s$, the evolution of the deflection is entirely governed by the memory of the GW burst. As the burst passes over Earth, a persistent deflection proportional to $m_0$ builds over the timescale $\tau$. The secular evolution of $|\delta n^i|$ between $t=0$ and $t=t_L$ is driven by the star term. Memory waveforms have, in a sense, semi-infinite wavelengths, so our previous argument for why the star term is negligible when the wavelength is small compared to $d_s$ does not apply. Though the star term is initially suppressed by the presence of $d_s$ in the denominator, the numerator of the star term, depending on the antiderivative of the GW waveform, grows linearly with time between $t=0$ and $t=t_L$ for memory. Since ${\cal V}^i_\oplus\neq{\cal V}^i_\bigstar$, the star term never fully cancels the prompt deflection generated through the Earth term. The most cancellation between the Earth and star terms occurs when $t\approx t_L/2$, where $|\delta n^i|$ can be seen in Fig. \ref{fig:deflection_norms} to go through a local minimum. Remember that $t_L$ is latitude dependent, explaining why the different colored curves go through minima at different times. If we increased the distance to our light sources, $d_s$, the vertical scaling of Fig. \ref{fig:deflection_norms} would not change. What would change is the horizontal span of the plot and the rate of secular evolution of $|\delta n^i|$ between $t=0$ and $t=t_L$. In other words, the memory-induced drift in the deflection driven by the star term is slower for more distant sources.

Instead of detailing the full evolution of the deflections from our representative GW burst, Fig. \ref{fig:displacements} shows the spatial variation across the sky for memory-induced deflections. Sources of light initially in the directions of the black circles will be promptly deflected to the positions of the red squares on a timescale $\tau$ as the memory component of the GW burst passes over Earth (we are ignoring the transient fluctuations from the waveletlike component of the burst). For the sake of visualization, the scale of the deflections has been enormously exaggerated. Over the timescale $(d_s/c)(1+{\bf\hat{z}\cdot\hat{n}})$, shorter for sources of light nearer the south pole, the promptly deflected sources will drift along the dotted green lines at a constant rate and eventually settle permanently at the locations of the blue triangles.

\paragraph{Discussion.---}
Light rays propagating through an inhomogeneous background of static, scalar density fluctuations undergo a random walk. The path they take deviates from the path they would take in a flat background by an amount that scales as $d_s^{3/2}$; their observed location is deflected by an angle that scales as $d_s^{1/2}$. It was initially thought that the story would be similar for deflections from a stochastic background of GWs. These expectations were thwarted, however, because the star term is suppressed, scaling as $d_s^{-1}$. Consequently, light rays propagating through a background of short-wavelength GWs (short compared to $d_s$) deviate from their unperturbed path by an amount that scales only logarithmically with $d_s$ and the deflection angle is independent of $d_s$ \cite{l86,kj97}.

Memory should revive random-walk scaling considerations for deflections from GWs. Well after a GW burst with memory has passed over Earth, the magnitude of the angular deflection of a distant light source is set by the scale of the memory rather than something scaling inversely with $d_s$. Populations of systems such as binary black holes---from stellar mass to supermassive---produce GW bursts with memory at some rate over a wide range of amplitudes. During the increased time that light takes to travel to Earth from more distant sources, more GW bursts with memory will occur and the persistent deflections from memory will partially stack in a random-walk fashion. That the deflection angle from memory scales precisely as $d_s^{1/2}$ rather than some other power of $d_s$ remains to be shown---secular evolution of the prompt deflection through the star term may complicate this picture slightly---but will be investigated in future work. Understanding how the magnitude of memory-induced deflections \emph{scales} with $d_s$ is one thing, but to estimate the \emph{magnitude} of the deflections will require astrophysically motivated models for the history of memory production from all possible sources reaching back over cosmological time. Furthermore, when considering the deflection signal from an ensemble of sources in different directions, one will need to look at how the spatial properties of the signature vary across the sky. It will be more complex than the single-source picture we developed here.

Even if the magnitude of memory-induced deflections builds to appreciable levels for distant sources in a random-walk fashion, these deflections may prove difficult to detect due to the simplistic nature of their temporal evolution. During the potentially very long interregnum between $t=0$ and $t=t_L$, a burst with memory of amplitude $m_0$ generates a constant proper motion field with a magnitude proportional to $cm_0/d_s$. Because it is the star term that drives the evolution of the deflection signal, it is in the magnitude of the memory-induced proper motion that the suppressive inverse factor of $d_s$ enters. For distant sources, these proper motions will likely be very small and the deflection signal will be nearly static over timescales comparable to a human lifetime. 

When tasked with looking for nearly static effects generated by GWs, it is useful to investigate entities for which there are well-motivated \emph{a priori} ideas for how they should look in the absence of said GWs, then considering the character and detectability of the GW imprint on them. This basic idea motivates searches for evidence of primordial gravitational waves in the cosmic microwave background (CMB) \cite{BICEP2+18}. As deflections from many memory events accumulate in a random-walk fashion, growing ever larger on average with increasing $d_s$, the largest possible effect will lie in the CMB. Describing the observational character of this signature is left to future work.

Our treatment of memory-induced deflections becomes increasingly unreliable as $d_s$ grows because for larger values of $d_s$, it is increasingly likely that bright GW bursts with memory occur at distances less than $d_s$. In these situations, it is inadequate to treat the GW wavefront as planar. \citet{bf11}, whose formalism we used for this analysis, also developed the machinery necessary to handle spherical wavefronts. It adds significant complexity to the problem. From experience developed through similar considerations applied to PTA efforts to detect GW memory \cite{mcc17}, we anticipate that the qualitative picture we have developed here will remain intact and that considerations involving spherical wavefronts will only become significant when considering lines of sight passing very close to bright sources of memory. 

Nonetheless, if one wishes to describe signatures of memory in the CMB, considerations pertaining to spherical wavefronts will be important. Further, one will need to impose the types of deflections we have described here to synthetic realizations of the CMB, but from model populations of the entire history of GW memory events that have occurred since recombination: stellar mass compact binary mergers, supermassive black hole binary mergers, etc. One will also need to investigate the extent to which GW memory-induced signatures in the CMB can be distinguished from or masked by, say, the distortions caused by gravitational lensing from large-scale structure in the foreground or the imprint of primordial GWs themselves. It is a difficult problem, rich with astrophysics, and worthy of exploration.

\vspace{.2cm}
D. R. M. is supported through a NANOGrav Physics Frontiers Center Postdoctoral Fellowship from the National Science Foundation Physics Frontiers Center Award No.~1430284.

\bibliographystyle{apsrev4-1}
\bibliography{relic.bib}

\begin{thebibliography}{32}%
\makeatletter
\providecommand \@ifxundefined [1]{%
 \@ifx{#1\undefined}
}%
\providecommand \@ifnum [1]{%
 \ifnum #1\expandafter \@firstoftwo
 \else \expandafter \@secondoftwo
 \fi
}%
\providecommand \@ifx [1]{%
 \ifx #1\expandafter \@firstoftwo
 \else \expandafter \@secondoftwo
 \fi
}%
\providecommand \natexlab [1]{#1}%
\providecommand \enquote  [1]{``#1''}%
\providecommand \bibnamefont  [1]{#1}%
\providecommand \bibfnamefont [1]{#1}%
\providecommand \citenamefont [1]{#1}%
\providecommand \href@noop [0]{\@secondoftwo}%
\providecommand \href [0]{\begingroup \@sanitize@url \@href}%
\providecommand \@href[1]{\@@startlink{#1}\@@href}%
\providecommand \@@href[1]{\endgroup#1\@@endlink}%
\providecommand \@sanitize@url [0]{\catcode `\\12\catcode `\$12\catcode
  `\&12\catcode `\#12\catcode `\^12\catcode `\_12\catcode `\%12\relax}%
\providecommand \@@startlink[1]{}%
\providecommand \@@endlink[0]{}%
\providecommand \url  [0]{\begingroup\@sanitize@url \@url }%
\providecommand \@url [1]{\endgroup\@href {#1}{\urlprefix }}%
\providecommand \urlprefix  [0]{URL }%
\providecommand \Eprint [0]{\href }%
\providecommand \doibase [0]{http://dx.doi.org/}%
\providecommand \selectlanguage [0]{\@gobble}%
\providecommand \bibinfo  [0]{\@secondoftwo}%
\providecommand \bibfield  [0]{\@secondoftwo}%
\providecommand \translation [1]{[#1]}%
\providecommand \BibitemOpen [0]{}%
\providecommand \bibitemStop [0]{}%
\providecommand \bibitemNoStop [0]{.\EOS\space}%
\providecommand \EOS [0]{\spacefactor3000\relax}%
\providecommand \BibitemShut  [1]{\csname bibitem#1\endcsname}%
\let\auto@bib@innerbib\@empty
\bibitem [{\citenamefont {{Braginskii}}\ and\ \citenamefont
  {{Thorne}}(1987)}]{bt87}%
  \BibitemOpen
  \bibfield  {author} {\bibinfo {author} {\bibfnamefont {V.~B.}\ \bibnamefont
  {{Braginskii}}}\ and\ \bibinfo {author} {\bibfnamefont {K.~S.}\ \bibnamefont
  {{Thorne}}},\ }\href {\doibase 10.1038/327123a0} {\bibfield  {journal}
  {\bibinfo  {journal} {\nat}\ }\textbf {\bibinfo {volume} {327}},\ \bibinfo
  {pages} {123} (\bibinfo {year} {1987})}\BibitemShut {NoStop}%
\bibitem [{\citenamefont {{Christodoulou}}(1991)}]{c91}%
  \BibitemOpen
  \bibfield  {author} {\bibinfo {author} {\bibfnamefont {D.}~\bibnamefont
  {{Christodoulou}}},\ }\href {\doibase 10.1103/PhysRevLett.67.1486} {\bibfield
   {journal} {\bibinfo  {journal} {\prl}\ }\textbf {\bibinfo {volume} {67}},\
  \bibinfo {pages} {1486} (\bibinfo {year} {1991})}\BibitemShut {NoStop}%
\bibitem [{\citenamefont {{Thorne}}(1992)}]{t92}%
  \BibitemOpen
  \bibfield  {author} {\bibinfo {author} {\bibfnamefont {K.~S.}\ \bibnamefont
  {{Thorne}}},\ }\href {\doibase 10.1103/PhysRevD.45.520} {\bibfield  {journal}
  {\bibinfo  {journal} {\prd}\ }\textbf {\bibinfo {volume} {45}},\ \bibinfo
  {pages} {520} (\bibinfo {year} {1992})}\BibitemShut {NoStop}%
\bibitem [{\citenamefont {{Blanchet}}\ and\ \citenamefont
  {{Damour}}(1992)}]{bd92}%
  \BibitemOpen
  \bibfield  {author} {\bibinfo {author} {\bibfnamefont {L.}~\bibnamefont
  {{Blanchet}}}\ and\ \bibinfo {author} {\bibfnamefont {T.}~\bibnamefont
  {{Damour}}},\ }\href {\doibase 10.1103/PhysRevD.46.4304} {\bibfield
  {journal} {\bibinfo  {journal} {\prd}\ }\textbf {\bibinfo {volume} {46}},\
  \bibinfo {pages} {4304} (\bibinfo {year} {1992})}\BibitemShut {NoStop}%
\bibitem [{\citenamefont {{Favata}}(2009)}]{f09}%
  \BibitemOpen
  \bibfield  {author} {\bibinfo {author} {\bibfnamefont {M.}~\bibnamefont
  {{Favata}}},\ }\href {\doibase 10.1088/0004-637X/696/2/L159} {\bibfield
  {journal} {\bibinfo  {journal} {Astrophys. J. Lett.}\ }\textbf {\bibinfo
  {volume} {696}},\ \bibinfo {pages} {L159} (\bibinfo {year} {2009})},\ \Eprint
  {http://arxiv.org/abs/0902.3660} {arXiv:0902.3660 [astro-ph.SR]} \BibitemShut
  {NoStop}%
\bibitem [{\citenamefont {{Favata}}(2010)}]{f10}%
  \BibitemOpen
  \bibfield  {author} {\bibinfo {author} {\bibfnamefont {M.}~\bibnamefont
  {{Favata}}},\ }\href {\doibase 10.1088/0264-9381/27/8/084036} {\bibfield
  {journal} {\bibinfo  {journal} {Classical and Quantum Gravity}\ }\textbf
  {\bibinfo {volume} {27}},\ \bibinfo {eid} {084036} (\bibinfo {year}
  {2010})},\ \Eprint {http://arxiv.org/abs/1003.3486} {arXiv:1003.3486 [gr-qc]}
  \BibitemShut {NoStop}%
\bibitem [{\citenamefont {{Talbot}}\ \emph {et~al.}(2018)\citenamefont
  {{Talbot}}, \citenamefont {{Thrane}}, \citenamefont {{Lasky}},\ and\
  \citenamefont {{Lin}}}]{ttl+18}%
  \BibitemOpen
  \bibfield  {author} {\bibinfo {author} {\bibfnamefont {C.}~\bibnamefont
  {{Talbot}}}, \bibinfo {author} {\bibfnamefont {E.}~\bibnamefont {{Thrane}}},
  \bibinfo {author} {\bibfnamefont {P.~D.}\ \bibnamefont {{Lasky}}}, \ and\
  \bibinfo {author} {\bibfnamefont {F.}~\bibnamefont {{Lin}}},\ }\href
  {\doibase 10.1103/PhysRevD.98.064031} {\bibfield  {journal} {\bibinfo
  {journal} {\prd}\ }\textbf {\bibinfo {volume} {98}},\ \bibinfo {eid} {064031}
  (\bibinfo {year} {2018})},\ \Eprint {http://arxiv.org/abs/1807.00990}
  {arXiv:1807.00990 [astro-ph.HE]} \BibitemShut {NoStop}%
\bibitem [{\citenamefont {Strominger}\ and\ \citenamefont
  {Zhiboedov}(2016)}]{sz16}%
  \BibitemOpen
  \bibfield  {author} {\bibinfo {author} {\bibfnamefont {A.}~\bibnamefont
  {Strominger}}\ and\ \bibinfo {author} {\bibfnamefont {A.}~\bibnamefont
  {Zhiboedov}},\ }\href {\doibase 10.1007/JHEP01(2016)086} {\bibfield
  {journal} {\bibinfo  {journal} {Journal of High Energy Physics}\ }\textbf
  {\bibinfo {volume} {2016}},\ \bibinfo {pages} {86} (\bibinfo {year}
  {2016})}\BibitemShut {NoStop}%
\bibitem [{\citenamefont {{Hawking}}\ \emph {et~al.}(2016)\citenamefont
  {{Hawking}}, \citenamefont {{Perry}},\ and\ \citenamefont
  {{Strominger}}}]{hps16}%
  \BibitemOpen
  \bibfield  {author} {\bibinfo {author} {\bibfnamefont {S.~W.}\ \bibnamefont
  {{Hawking}}}, \bibinfo {author} {\bibfnamefont {M.~J.}\ \bibnamefont
  {{Perry}}}, \ and\ \bibinfo {author} {\bibfnamefont {A.}~\bibnamefont
  {{Strominger}}},\ }\href {\doibase 10.1103/PhysRevLett.116.231301} {\bibfield
   {journal} {\bibinfo  {journal} {\prl}\ }\textbf {\bibinfo {volume} {116}},\
  \bibinfo {eid} {231301} (\bibinfo {year} {2016})},\ \Eprint
  {http://arxiv.org/abs/1601.00921} {arXiv:1601.00921 [hep-th]} \BibitemShut
  {NoStop}%
\bibitem [{\citenamefont {{Hawking}}\ \emph {et~al.}(2017)\citenamefont
  {{Hawking}}, \citenamefont {{Perry}},\ and\ \citenamefont
  {{Strominger}}}]{hps17}%
  \BibitemOpen
  \bibfield  {author} {\bibinfo {author} {\bibfnamefont {S.~W.}\ \bibnamefont
  {{Hawking}}}, \bibinfo {author} {\bibfnamefont {M.~J.}\ \bibnamefont
  {{Perry}}}, \ and\ \bibinfo {author} {\bibfnamefont {A.}~\bibnamefont
  {{Strominger}}},\ }\href {\doibase 10.1007/JHEP05(2017)161} {\bibfield
  {journal} {\bibinfo  {journal} {Journal of High Energy Physics}\ }\textbf
  {\bibinfo {volume} {2017}},\ \bibinfo {eid} {161} (\bibinfo {year} {2017})},\
  \Eprint {http://arxiv.org/abs/1611.09175} {arXiv:1611.09175 [hep-th]}
  \BibitemShut {NoStop}%
\bibitem [{\citenamefont {{Hollands}}\ \emph {et~al.}(2017)\citenamefont
  {{Hollands}}, \citenamefont {{Ishibashi}},\ and\ \citenamefont
  {{Wald}}}]{hiw17}%
  \BibitemOpen
  \bibfield  {author} {\bibinfo {author} {\bibfnamefont {S.}~\bibnamefont
  {{Hollands}}}, \bibinfo {author} {\bibfnamefont {A.}~\bibnamefont
  {{Ishibashi}}}, \ and\ \bibinfo {author} {\bibfnamefont {R.~M.}\ \bibnamefont
  {{Wald}}},\ }\href {\doibase 10.1088/1361-6382/aa777a} {\bibfield  {journal}
  {\bibinfo  {journal} {Classical and Quantum Gravity}\ }\textbf {\bibinfo
  {volume} {34}},\ \bibinfo {eid} {155005} (\bibinfo {year} {2017})},\ \Eprint
  {http://arxiv.org/abs/1612.03290} {arXiv:1612.03290 [gr-qc]} \BibitemShut
  {NoStop}%
\bibitem [{\citenamefont {{Flanagan}}\ and\ \citenamefont
  {{Nichols}}(2017)}]{fn17}%
  \BibitemOpen
  \bibfield  {author} {\bibinfo {author} {\bibfnamefont {{\'E}.~{\'E}.}\
  \bibnamefont {{Flanagan}}}\ and\ \bibinfo {author} {\bibfnamefont {D.~A.}\
  \bibnamefont {{Nichols}}},\ }\href {\doibase 10.1103/PhysRevD.95.044002}
  {\bibfield  {journal} {\bibinfo  {journal} {\prd}\ }\textbf {\bibinfo
  {volume} {95}},\ \bibinfo {eid} {044002} (\bibinfo {year} {2017})},\ \Eprint
  {http://arxiv.org/abs/1510.03386} {arXiv:1510.03386 [hep-th]} \BibitemShut
  {NoStop}%
\bibitem [{\citenamefont {{van Haasteren}}\ and\ \citenamefont
  {{Levin}}(2010)}]{vl10}%
  \BibitemOpen
  \bibfield  {author} {\bibinfo {author} {\bibfnamefont {R.}~\bibnamefont {{van
  Haasteren}}}\ and\ \bibinfo {author} {\bibfnamefont {Y.}~\bibnamefont
  {{Levin}}},\ }\href {\doibase 10.1111/j.1365-2966.2009.15885.x} {\bibfield
  {journal} {\bibinfo  {journal} {MNRAS}\ }\textbf {\bibinfo {volume} {401}},\
  \bibinfo {pages} {2372} (\bibinfo {year} {2010})},\ \Eprint
  {http://arxiv.org/abs/0909.0954} {arXiv:0909.0954 [astro-ph.IM]} \BibitemShut
  {NoStop}%
\bibitem [{\citenamefont {{Cordes}}\ and\ \citenamefont
  {{Jenet}}(2012)}]{cj12}%
  \BibitemOpen
  \bibfield  {author} {\bibinfo {author} {\bibfnamefont {J.~M.}\ \bibnamefont
  {{Cordes}}}\ and\ \bibinfo {author} {\bibfnamefont {F.~A.}\ \bibnamefont
  {{Jenet}}},\ }\href {\doibase 10.1088/0004-637X/752/1/54} {\bibfield
  {journal} {\bibinfo  {journal} {\apj}\ }\textbf {\bibinfo {volume} {752}},\
  \bibinfo {eid} {54} (\bibinfo {year} {2012})}\BibitemShut {NoStop}%
\bibitem [{\citenamefont {{Madison}}\ \emph {et~al.}(2014)\citenamefont
  {{Madison}}, \citenamefont {{Cordes}},\ and\ \citenamefont
  {{Chatterjee}}}]{mcc14}%
  \BibitemOpen
  \bibfield  {author} {\bibinfo {author} {\bibfnamefont {D.~R.}\ \bibnamefont
  {{Madison}}}, \bibinfo {author} {\bibfnamefont {J.~M.}\ \bibnamefont
  {{Cordes}}}, \ and\ \bibinfo {author} {\bibfnamefont {S.}~\bibnamefont
  {{Chatterjee}}},\ }\href {\doibase 10.1088/0004-637X/788/2/141} {\bibfield
  {journal} {\bibinfo  {journal} {\apj}\ }\textbf {\bibinfo {volume} {788}},\
  \bibinfo {eid} {141} (\bibinfo {year} {2014})},\ \Eprint
  {http://arxiv.org/abs/1404.5682} {arXiv:1404.5682 [astro-ph.HE]} \BibitemShut
  {NoStop}%
\bibitem [{\citenamefont {{Wang}}\ \emph {et~al.}(2015)\citenamefont {{Wang}}
  \emph {et~al.}}]{whc+15}%
  \BibitemOpen
  \bibfield  {author} {\bibinfo {author} {\bibfnamefont {J.~B.}\ \bibnamefont
  {{Wang}}} \emph {et~al.},\ }\href {\doibase 10.1093/mnras/stu2137} {\bibfield
   {journal} {\bibinfo  {journal} {MNRAS}\ }\textbf {\bibinfo {volume} {446}},\
  \bibinfo {pages} {1657} (\bibinfo {year} {2015})},\ \Eprint
  {http://arxiv.org/abs/1410.3323} {arXiv:1410.3323 [astro-ph.GA]} \BibitemShut
  {NoStop}%
\bibitem [{\citenamefont {{Aggarwal}}\ \emph {et~al.}(2020)\citenamefont
  {{Aggarwal}} \emph {et~al.}}]{aab+20}%
  \BibitemOpen
  \bibfield  {author} {\bibinfo {author} {\bibfnamefont {K.}~\bibnamefont
  {{Aggarwal}}} \emph {et~al.},\ }\href {\doibase 10.3847/1538-4357/ab6083}
  {\bibfield  {journal} {\bibinfo  {journal} {\apj}\ }\textbf {\bibinfo
  {volume} {889}},\ \bibinfo {eid} {38} (\bibinfo {year} {2020})},\ \Eprint
  {http://arxiv.org/abs/1911.08488} {arXiv:1911.08488 [astro-ph.HE]}
  \BibitemShut {NoStop}%
\bibitem [{\citenamefont {{Islo}}\ \emph {et~al.}(2019)\citenamefont {{Islo}},
  \citenamefont {{Simon}}, \citenamefont {{Burke-Spolaor}},\ and\ \citenamefont
  {{Siemens}}}]{isb+19}%
  \BibitemOpen
  \bibfield  {author} {\bibinfo {author} {\bibfnamefont {K.}~\bibnamefont
  {{Islo}}}, \bibinfo {author} {\bibfnamefont {J.}~\bibnamefont {{Simon}}},
  \bibinfo {author} {\bibfnamefont {S.}~\bibnamefont {{Burke-Spolaor}}}, \ and\
  \bibinfo {author} {\bibfnamefont {X.}~\bibnamefont {{Siemens}}},\ }\href@noop
  {} {\bibfield  {journal} {\bibinfo  {journal} {arXiv e-prints}\ ,\ \bibinfo
  {eid} {arXiv:1906.11936}} (\bibinfo {year} {2019})},\ \Eprint
  {http://arxiv.org/abs/1906.11936} {arXiv:1906.11936 [astro-ph.HE]}
  \BibitemShut {NoStop}%
\bibitem [{\citenamefont {{Lasky}}\ \emph {et~al.}(2016)\citenamefont {{Lasky}}
  \emph {et~al.}}]{ltl+16}%
  \BibitemOpen
  \bibfield  {author} {\bibinfo {author} {\bibfnamefont {P.~D.}\ \bibnamefont
  {{Lasky}}} \emph {et~al.},\ }\href {\doibase 10.1103/PhysRevLett.117.061102}
  {\bibfield  {journal} {\bibinfo  {journal} {\prl}\ }\textbf {\bibinfo
  {volume} {117}},\ \bibinfo {eid} {061102} (\bibinfo {year} {2016})},\ \Eprint
  {http://arxiv.org/abs/1605.01415} {arXiv:1605.01415 [astro-ph.HE]}
  \BibitemShut {NoStop}%
\bibitem [{\citenamefont {{H{\"u}bner}}\ \emph {et~al.}(2020)\citenamefont
  {{H{\"u}bner}}, \citenamefont {{Talbot}}, \citenamefont {{Lasky}},\ and\
  \citenamefont {{Thrane}}}]{htl+20}%
  \BibitemOpen
  \bibfield  {author} {\bibinfo {author} {\bibfnamefont {M.}~\bibnamefont
  {{H{\"u}bner}}}, \bibinfo {author} {\bibfnamefont {C.}~\bibnamefont
  {{Talbot}}}, \bibinfo {author} {\bibfnamefont {P.~D.}\ \bibnamefont
  {{Lasky}}}, \ and\ \bibinfo {author} {\bibfnamefont {E.}~\bibnamefont
  {{Thrane}}},\ }\href {\doibase 10.1103/PhysRevD.101.023011} {\bibfield
  {journal} {\bibinfo  {journal} {\prd}\ }\textbf {\bibinfo {volume} {101}},\
  \bibinfo {eid} {023011} (\bibinfo {year} {2020})},\ \Eprint
  {http://arxiv.org/abs/1911.12496} {arXiv:1911.12496 [astro-ph.HE]}
  \BibitemShut {NoStop}%
\bibitem [{\citenamefont {{Kopeikin}}\ \emph {et~al.}(1999)\citenamefont
  {{Kopeikin}}, \citenamefont {{Sch{\"a}fer}}, \citenamefont {{Gwinn}},\ and\
  \citenamefont {{Eubanks}}}]{ksg+99}%
  \BibitemOpen
  \bibfield  {author} {\bibinfo {author} {\bibfnamefont {S.~M.}\ \bibnamefont
  {{Kopeikin}}}, \bibinfo {author} {\bibfnamefont {G.}~\bibnamefont
  {{Sch{\"a}fer}}}, \bibinfo {author} {\bibfnamefont {C.~R.}\ \bibnamefont
  {{Gwinn}}}, \ and\ \bibinfo {author} {\bibfnamefont {T.~M.}\ \bibnamefont
  {{Eubanks}}},\ }\href {\doibase 10.1103/PhysRevD.59.084023} {\bibfield
  {journal} {\bibinfo  {journal} {\prd}\ }\textbf {\bibinfo {volume} {59}},\
  \bibinfo {eid} {084023} (\bibinfo {year} {1999})},\ \Eprint
  {http://arxiv.org/abs/gr-qc/9811003} {arXiv:gr-qc/9811003 [gr-qc]}
  \BibitemShut {NoStop}%
\bibitem [{\citenamefont {{Book}}\ and\ \citenamefont
  {{Flanagan}}(2011)}]{bf11}%
  \BibitemOpen
  \bibfield  {author} {\bibinfo {author} {\bibfnamefont {L.~G.}\ \bibnamefont
  {{Book}}}\ and\ \bibinfo {author} {\bibfnamefont {{\'E}.~{\'E}.}\
  \bibnamefont {{Flanagan}}},\ }\href {\doibase 10.1103/PhysRevD.83.024024}
  {\bibfield  {journal} {\bibinfo  {journal} {\prd}\ }\textbf {\bibinfo
  {volume} {83}},\ \bibinfo {eid} {024024} (\bibinfo {year} {2011})},\ \Eprint
  {http://arxiv.org/abs/1009.4192} {arXiv:1009.4192 [astro-ph.CO]} \BibitemShut
  {NoStop}%
\bibitem [{\citenamefont {{Schutz}}(2009)}]{s09}%
  \BibitemOpen
  \bibfield  {author} {\bibinfo {author} {\bibfnamefont {B.~F.}\ \bibnamefont
  {{Schutz}}},\ }in\ \href@noop {} {\emph {\bibinfo {booktitle} {IAU Symposium
  \#261, American Astronomical Society}}},\ Vol.~\bibinfo {volume} {41}\
  (\bibinfo {year} {2009})\ p.\ \bibinfo {pages} {888}\BibitemShut {NoStop}%
\bibitem [{\citenamefont {{Gaia Collaboration}}(2018)}]{Gaia18}%
  \BibitemOpen
  \bibfield  {author} {\bibinfo {author} {\bibnamefont {{Gaia
  Collaboration}}},\ }\href {\doibase 10.1051/0004-6361/201833051} {\bibfield
  {journal} {\bibinfo  {journal} {A\&A}\ }\textbf {\bibinfo {volume} {616}},\
  \bibinfo {eid} {A1} (\bibinfo {year} {2018})},\ \Eprint
  {http://arxiv.org/abs/1804.09365} {arXiv:1804.09365 [astro-ph.GA]}
  \BibitemShut {NoStop}%
\bibitem [{\citenamefont {{Moore}}\ \emph {et~al.}(2017)\citenamefont
  {{Moore}}, \citenamefont {{Mihaylov}}, \citenamefont {{Lasenby}},\ and\
  \citenamefont {{Gilmore}}}]{mml+17}%
  \BibitemOpen
  \bibfield  {author} {\bibinfo {author} {\bibfnamefont {C.~J.}\ \bibnamefont
  {{Moore}}}, \bibinfo {author} {\bibfnamefont {D.~P.}\ \bibnamefont
  {{Mihaylov}}}, \bibinfo {author} {\bibfnamefont {A.}~\bibnamefont
  {{Lasenby}}}, \ and\ \bibinfo {author} {\bibfnamefont {G.}~\bibnamefont
  {{Gilmore}}},\ }\href {\doibase 10.1103/PhysRevLett.119.261102} {\bibfield
  {journal} {\bibinfo  {journal} {\prl}\ }\textbf {\bibinfo {volume} {119}},\
  \bibinfo {eid} {261102} (\bibinfo {year} {2017})},\ \Eprint
  {http://arxiv.org/abs/1707.06239} {arXiv:1707.06239 [astro-ph.IM]}
  \BibitemShut {NoStop}%
\bibitem [{\citenamefont {{Event Horizon Telescope Collaboration}}\ \emph
  {et~al.}(2019)\citenamefont {{Event Horizon Telescope Collaboration}} \emph
  {et~al.}}]{EHT19}%
  \BibitemOpen
  \bibfield  {author} {\bibinfo {author} {\bibnamefont {{Event Horizon
  Telescope Collaboration}}} \emph {et~al.},\ }\href {\doibase
  10.3847/2041-8213/ab0ec7} {\bibfield  {journal} {\bibinfo  {journal} {\apj}\
  }\textbf {\bibinfo {volume} {875}},\ \bibinfo {eid} {L1} (\bibinfo {year}
  {2019})},\ \Eprint {http://arxiv.org/abs/1906.11238} {arXiv:1906.11238
  [astro-ph.GA]} \BibitemShut {NoStop}%
\bibitem [{\citenamefont {{Darling}}\ \emph {et~al.}(2018)\citenamefont
  {{Darling}}, \citenamefont {{Truebenbach}},\ and\ \citenamefont
  {{Paine}}}]{dtp18}%
  \BibitemOpen
  \bibfield  {author} {\bibinfo {author} {\bibfnamefont {J.}~\bibnamefont
  {{Darling}}}, \bibinfo {author} {\bibfnamefont {A.~E.}\ \bibnamefont
  {{Truebenbach}}}, \ and\ \bibinfo {author} {\bibfnamefont {J.}~\bibnamefont
  {{Paine}}},\ }\href {\doibase 10.3847/1538-4357/aac772} {\bibfield  {journal}
  {\bibinfo  {journal} {\apj}\ }\textbf {\bibinfo {volume} {861}},\ \bibinfo
  {eid} {113} (\bibinfo {year} {2018})},\ \Eprint
  {http://arxiv.org/abs/1804.06986} {arXiv:1804.06986 [astro-ph.IM]}
  \BibitemShut {NoStop}%
\bibitem [{\citenamefont {{Pyne}}\ \emph {et~al.}(1996)\citenamefont {{Pyne}}
  \emph {et~al.}}]{pgb+96}%
  \BibitemOpen
  \bibfield  {author} {\bibinfo {author} {\bibfnamefont {T.}~\bibnamefont
  {{Pyne}}} \emph {et~al.},\ }\href {\doibase 10.1086/177443} {\bibfield
  {journal} {\bibinfo  {journal} {\apj}\ }\textbf {\bibinfo {volume} {465}},\
  \bibinfo {pages} {566} (\bibinfo {year} {1996})},\ \Eprint
  {http://arxiv.org/abs/astro-ph/9507030} {arXiv:astro-ph/9507030 [astro-ph]}
  \BibitemShut {NoStop}%
\bibitem [{\citenamefont {{Linder}}(1986)}]{l86}%
  \BibitemOpen
  \bibfield  {author} {\bibinfo {author} {\bibfnamefont {E.~V.}\ \bibnamefont
  {{Linder}}},\ }\href {\doibase 10.1103/PhysRevD.34.1759} {\bibfield
  {journal} {\bibinfo  {journal} {\prd}\ }\textbf {\bibinfo {volume} {34}},\
  \bibinfo {pages} {1759} (\bibinfo {year} {1986})}\BibitemShut {NoStop}%
\bibitem [{\citenamefont {{Kaiser}}\ and\ \citenamefont
  {{Jaffe}}(1997)}]{kj97}%
  \BibitemOpen
  \bibfield  {author} {\bibinfo {author} {\bibfnamefont {N.}~\bibnamefont
  {{Kaiser}}}\ and\ \bibinfo {author} {\bibfnamefont {A.}~\bibnamefont
  {{Jaffe}}},\ }\href {\doibase 10.1086/304357} {\bibfield  {journal} {\bibinfo
   {journal} {\apj}\ }\textbf {\bibinfo {volume} {484}},\ \bibinfo {pages}
  {545} (\bibinfo {year} {1997})},\ \Eprint
  {http://arxiv.org/abs/astro-ph/9609043} {arXiv:astro-ph/9609043 [astro-ph]}
  \BibitemShut {NoStop}%
\bibitem [{\citenamefont {{BICEP2 Collaboration}}\ \emph
  {et~al.}(2018)\citenamefont {{BICEP2 Collaboration}}, \citenamefont {{Keck
  Array Collaboration}} \emph {et~al.}}]{BICEP2+18}%
  \BibitemOpen
  \bibfield  {author} {\bibinfo {author} {\bibnamefont {{BICEP2
  Collaboration}}}, \bibinfo {author} {\bibnamefont {{Keck Array
  Collaboration}}},  \emph {et~al.},\ }\href {\doibase
  10.1103/PhysRevLett.121.221301} {\bibfield  {journal} {\bibinfo  {journal}
  {\prl}\ }\textbf {\bibinfo {volume} {121}},\ \bibinfo {eid} {221301}
  (\bibinfo {year} {2018})},\ \Eprint {http://arxiv.org/abs/1810.05216}
  {arXiv:1810.05216 [astro-ph.CO]} \BibitemShut {NoStop}%
\bibitem [{\citenamefont {{Madison}}\ \emph {et~al.}(2017)\citenamefont
  {{Madison}}, \citenamefont {{Chernoff}},\ and\ \citenamefont
  {{Cordes}}}]{mcc17}%
  \BibitemOpen
  \bibfield  {author} {\bibinfo {author} {\bibfnamefont {D.~R.}\ \bibnamefont
  {{Madison}}}, \bibinfo {author} {\bibfnamefont {D.~F.}\ \bibnamefont
  {{Chernoff}}}, \ and\ \bibinfo {author} {\bibfnamefont {J.~M.}\ \bibnamefont
  {{Cordes}}},\ }\href {\doibase 10.1103/PhysRevD.96.123016} {\bibfield
  {journal} {\bibinfo  {journal} {\prd}\ }\textbf {\bibinfo {volume} {96}},\
  \bibinfo {eid} {123016} (\bibinfo {year} {2017})},\ \Eprint
  {http://arxiv.org/abs/1710.04974} {arXiv:1710.04974 [astro-ph.GA]}
  \BibitemShut {NoStop}%
\end{thebibliography}%
\end{document}